\def\beq{\begin{equation}}
\def\eeq{\end{equation}}
\documentclass[twocolumn,showpacs,preprintnumbers,amsmath,amssymb]{revtex4}
\usepackage{graphicx}
\usepackage{dcolumn}
\usepackage{bm}
\begin{document}
\def\s{\rule{0in}{2.8ex}}

\title{Application of wavelets to singular integral scattering equations}

\author{B. M. Kessler}

\email{brian-kessler@uiowa.edu}
\author{G. L. Payne}
\author{W. N. Polyzou}
\affiliation{%
Department of Physics and Astronomy, \\The 
University of Iowa  \\
Iowa City, IA  52242 
}

\date{\today}

\begin{abstract}
The use of orthonormal wavelet basis functions for solving singular integral
scattering equations is investigated. It is shown that these basis functions
lead to sparse matrix equations which can be solved by iterative techniques.
The scaling properties of wavelets are used to derive an efficient method for
evaluating the singular integrals. The accuracy and efficiency of the 
wavelet transforms is demonstrated by solving the two-body T-matrix equation
without partial wave projection. The resulting matrix equation which is
characteristic of multiparticle integral scattering equations is found to
provide an efficient method for obtaining accurate approximate solutions to
the integral equation. These results indicate that wavelet transforms may
provide a useful tool for studying few-body systems.

\end{abstract}


\maketitle

\section{Introduction}

Few-body systems provide a useful tool for studying the dynamics
of hadronic systems. The combination of short-ranged interactions and
finite density means that the dynamics of complex hadronic systems can
be understood by studying the dynamics of few-degree of freedom
sub-systems. Few-body systems are simple enough to perform nearly
complete high-precision measurements and to perform ab-initio
calculations that are exact to within the experimental precision.
This clean connection between theory and experiment has led to an
excellent understanding of two-body interactions in
low-energy nuclear physics, and a good understanding of the three-body
interactions.

Our knowledge of low-energy hadronic dynamics is largely due to 
the interplay between
experimental and computational advances. A complete understanding of
even the simplest few-hadron system requires measurements of a
complete set of spin observables which have small cross sections and
require state of the art detectors. At the same time, the model
calculations with realistic interactions are limited by computer speed
and memory. In addition the equations are either singular or have
complicated boundary conditions which require specialized numerical
treatments.

One of the most interesting energy scales is the one where the
natural choice of few-body degrees of freedom changes from nucleons
and mesons to sub-nucleon degrees of freedom. The QCD string tension
or nucleon size suggest that the relevant scale for the onset of this
transition is about a GeV. A consistent dynamics of hadrons or
sub-nuclear particles on this scale must be relativistic; a Galilean
invariant theory cannot simultaneously preserve momentum
conservation in the lab and center of momentum frames if the initial
and final reaction products have different masses. Relativistic dynamical
 models are most naturally formulated in
momentum space. This is due to the presence of momentum-dependent
Wigner and/or Melosh rotations as well as square roots that appear in
the relationship between energy and momentum.

Non-relativistic few-body calculations formulated in configuration
space with local potentials have the advantage that the matrices
obtained after discretizing the dynamical equations are banded, thus
reducing the size of the numerical calculations. Equivalent
momentum-space calculations lead to dense matrices of comparable dimensions.
In addition, the embedding of the two-body interactions in the
three-body Hilbert space leads to non-localities. Realistic
relativistic three-body calculations are just beginning to be solved
\cite{stadler,glockle}. Numerical methods that can reduce the
size of these calculations could make relativistic calculations of
realistic systems more tractable.

In this paper we explore the use of wavelet basis
functions to reduce the size of momentum space scattering calculations. The resulting linear system can be can be accurately
approximated by a linear system with a sparse kernel. It is our
contention that the use of this sparse kernel results in a reduction
in the size of the numerical calculation that is comparable to the
corresponding configuration space calculations. The advantage is that
the wavelet methods can be applied in momentum space and are not
limited to local interactions.

The long-term goal is to apply wavelet methods to solve the
relativistic three-body problem. In a previous paper \cite{fewbody}, we
tested this method to solve the non-relativistic Lippmann-Schwinger
equation with a Malfliet Tjon V potential. In this test problem,
the s-wave K-matrix was computed. The wavelet method led to a
significant reduction in the size of the problem. We found that 96\%
of the matrix elements of the kernel of the integral equation could be
eliminated leading to an error of only a few parts in a million.

The success of wavelet method in \cite{fewbody} suggests that the method
should be tested on a more complicated problem. In this paper, we test
the wavelet method on the same problem without using partial waves.
This leads to a singular two-variable integral equation, which has the
same number of continuous variables as the three-body Faddeev
equations with partial waves. It is simpler than the full three-body
calculation, but is a much larger calculation than was needed in Ref.
\cite{fewbody}. In addition, computations that employ conventional 
methods \cite{char1} are available for comparison. In solving this problem it is 
necessary to address issues involving the storage and computations 
with large matrices.
 
One well known use of
wavelets is in the data compression algorithm used in JPEG files \cite{jpeg}. 
Our motivation for applying wavelet methods to scattering problems is
based on the observation that both a digital photograph and a
discretized kernel of an integral equation are two-dimensional arrays
of numbers. If wavelets can reduce the size of a digital image, they
should have a similar effect on the size of the kernel of an integral
equation.

Given the utility of wavelets in digital data processing, it is natural
to ask why they have not been used extensively in numerical
computations in scattering. One possible reason is because there is a
non-trivial learning curve that must be overcome for a successful
application to singular integral equations. A relevant feature is
that the basis functions have a fractal structure; they are solutions
to a linear renormalization group equation and thus have structure on
all scales. Numerical techniques that exploit the local smoothness of
functions do not work effectively with functions that have structure
on all scales.

In \cite{fewbody}, we concluded that these limitations could be overcome by
exploiting the renormalization group transformation properties of the
basis functions in numerical computations. These equations were used
to compute moments of the basis functions with polynomials. These
moments were used to construct efficient quadrature methods for
evaluating overlap integrals. In addition, these moments could be
combined with the renormalization group equations to perform accurate
calculations of the types of singular integrals that appear in
scattering problems. A key conclusion of
\cite{fewbody} was that wavelet methods provide an accurate and effective 
method for solving the scattering equations. In addition, the 
expected reduction in the size of the numerical problem could be 
achieved with minimal loss of precision.

There are many kinds of wavelets. In \cite{fewbody} we found that the
Daubechies-3 \cite{daub} wavelets proved to be the most useful for our
calculations.  Numerical methods based on wavelets utilize the
existence of two orthogonal bases for a model space. The two bases are
related by an orthogonal transformation. The first basis, called the
father function basis, samples the data by averaging on small
scales. It is the numerical equivalent of a raw digital
photograph. The orthogonal transformation is generated by filtering
the coefficients of the father function basis into equal numbers of
high and low frequency parts. The high frequency parts are associated 
with another type of basis function known as the mother function.
   The same filter is again applied only to
the to the remaining low frequency parts, which are divided into high
and low frequency parts. This is repeated until there is only one low
frequency coefficient. This orthogonal transformation and its inverse
can be generated with the same type of efficiency as a fast Fourier
transform.  The new basis is called the wavelet basis.
  
For the Daubechies-3 wavelets, both sets of basis functions have
compact support. The support of the father function basis functions
is small and is determined by the resolution of the model space. The
support of the wavelet basis functions is compact, but occurs on all
scales between the finest resolution and the coarsest resolution.

The father function for the Daubechies-3 wavelets has the
property that a finite linear combination of such functions can
locally pointwise represent a polynomial of degree two or
less. Integrals over these polynomials and the scaling basis functions
can be done exactly and efficiently using a one-point quadrature.

The mother functions have the property that they are orthogonal
to polynomials of degree two. This means that the expansion
coefficient for a given mother basis function is zero if the function
can be well-approximated by a polynomial on the 
support of the basis
function. It is for this reason that most of the kernel matrix
elements in this representation are small. Setting these small
coefficients to zero is the key approximation that leads to 
sparse matrices.
 
Some of the properties that make the Daubechies wavelets interesting for 
numerical computations are 
\begin{itemize}
\item The basis functions have compact support.
\vspace{-4 pt}
\item The basis functions are orthonormal.
\vspace{-4 pt}
\item The basis functions can pointwise represent low degree polynomials
\vspace{-4 pt}
\item The wavelet transform automatically identifies the important 
basis functions.
\vspace{-4 pt}
\item There is a simple one point quadrature rule that is exact for 
low-degree local polynomials.
\vspace{-4 pt}
\item These are accurate methods for computing the singular integrals of 
scattering theory.
\vspace{-4 pt}
\item The basis functions never have to be computed. 
\end{itemize}

The above list indicates that wavelet bases have many advantages in
common with spline bases, which have proven to be very useful in large
few-body calculations.  Both the spline and wavelet basis functions
have compact support, which allows them to efficiently model local
structures, both provide pointwise representations of low-degree
polynomials, both can be easily integrated using simple quadrature
rules, and both can be accurately integrated over the scattering
singularity.  One feature that distinguishes the wavelet method from
the spline method is that the wavelet transform automatically
identifies the important basis functions that need to be retained.
With splines, the regions that have a lot of structure and require
extra splines need to be identified by hand.  This is a non-trivial
problem in large calculations.  The automatic nature of this step is
an important advantage of the wavelet method in large calculations.
In addition, unlike the spline basis functions, the wavelet basis
functions are orthogonal, and the one-point quadrature only requires
the evaluation of the driving term or kernel at a single point to
compute matrix elements.  This leads to numerical approximations 
that combine the efficiency of the collocation method with the 
stability of the Galerkin method.

In the next section we give an overview of the properties of wavelets
that are used in our numerical computations.  Our model problem is
defined in section three.  The methods of section two are used in
section four to reduce the scattering integral equation in section
three to an approximate linear system.  The transformation to a
sparse-matrix linear system and the methods used to solve the linear
equations are discussed in section five.  The considerations discussed
in this section are important for realistic applications.  The results
of the model calculations are discussed and compared to the results of
partial-wave calculations in section six.  Our conclusions are
summarized in section seven.  The complex biconjugate gradient
algorithm that was used to solve the resulting system of linear
equations is outlined in the Appendix.

\section{Wavelet Properties}

In our work, we use Daubechies' original bases of compactly supported wavelets \cite{daub}.  In addition to their simplicity, these functions possess many useful properties for numeric calculations, which are discussed at the end of this section.

\subsection{General Wavelet Analysis}

There are two primal basis functions called the father, $\phi$, and mother, $\psi$.  The primal father function is defined as the solution of the homogeneous scaling equation

\begin{equation}
\phi(x) = \sqrt{2} \sum_{l=0}^{2K-1} h_l \phi(2x-l),
\label{scale}
\end{equation}
with normalization
\begin{equation}
\int \phi(x) dx = 1.
\end{equation}
The primal mother function is defined in terms of the father by a similar scaling equation,
\begin{equation}
\psi(x) = \sqrt{2} \sum_{l=0}^{2K-1} g_l \phi(2x-l),
\end{equation}
where
\begin{equation}
g_l = (-1)^l h_{2K-1-l}.
\end{equation}
The parameter $K$ is the order of the Daubechies wavelet and the $h_l$ are a unique set of numerical coefficients that satisfy certain relations \cite{daub} such as orthogonality of basis functions.  We employ wavelets of order $K=3$, henceforth called Daubechies-3 wavelets.   The numerical values of the $h_l$ are given in Table \ref{coef}.

Equation (\ref{scale}) is the most important in all of wavelet analysis, as all the properties of a wavelet basis are determined by the so-called filter coefficients, $h_l$.  A simple property that follows from the $h_l$ is that the father and mother function both have compact support on the interval $(0,2K-1)$.  All other basis functions are related to the primal father and mother by means of dyadic (power of two) scale transformations and unit translations,

\begin{table}[t]
\caption{Scaling Coefficients for Daubechies-3 Wavelets}
\begin{tabular}{|l|l|}
\hline				      		      
\s$h_0$ & $(1+\sqrt{10}+\sqrt{5+2\sqrt{10}}\,)/16\sqrt{2}$ \\
$h_1$ & $(5+\sqrt{10}+3\sqrt{5+2\sqrt{10}}\,)/16\sqrt{2}$ \\
$h_2$ & $(10-2\sqrt{10}+2\sqrt{5+2\sqrt{10}}\,)/16\sqrt{2}$ \\
$h_3$ & $ (10-2\sqrt{10}-2\sqrt{5+2\sqrt{10}}\,)/16\sqrt{2} $ \\
$h_4$ & $(5+\sqrt{10}-3\sqrt{5+2\sqrt{10}}\,)/16\sqrt{2}$ \\
$h_5$ & $(1+\sqrt{10}-\sqrt{5+2\sqrt{10}}\,)/16\sqrt{2}$ \\
\hline
\end{tabular}
\label{coef}
\end{table}

\begin{eqnarray}
\phi_{j,k}(x) :=2^{-j/2}\phi(2^jx-k) \,\, && \nonumber \\ \psi_{j,k}(x):=2^{-j/2}\psi(2^jx-k). &&
\end{eqnarray}

\begin{figure}[h]
\begin{center}
\rotatebox{00}{\resizebox{3.5in}{!}{
\includegraphics{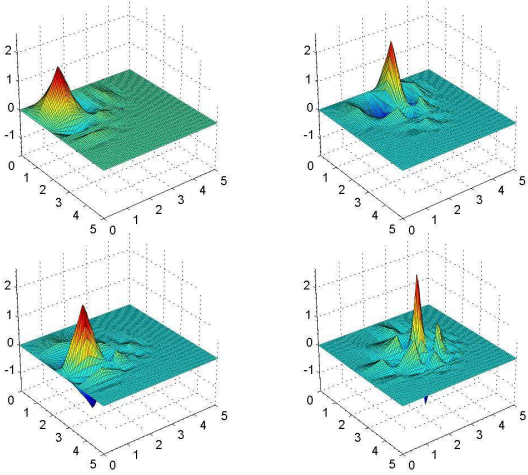}}
}
\caption{(Color online) Direct Product Basis of Daubechies-3 Wavelets}
\label{basis}
\end{center}
\end{figure}

To solve the two-dimensional integral equation for the T-matrix we need to construct a two-dimensional direct product basis consisting of functions of the form

\begin{eqnarray}
\phi_{m,l}(x)\phi_{n,k}(y), & \quad & \phi_{m,l}(x)\psi_{n,k}(y),\nonumber  \\ \psi_{m,l}(x)\phi_{n,k}(y), &\quad \mathrm{and} \quad & \psi_{m,l}(x)\psi_{n,k}(y).
\end{eqnarray}
The primal versions of these four basis function types for the Daubechies-3 wavelets are shown in Fig. \ref{basis}.

\subsection{Equivalent Representations and Wavelet Transforms}

If one includes wavelets of all scales, then one can obtain a basis for $L^2(\mathbb{R})$.  In practice however, one chooses a fine approximation scale $J$ and constructs an approximation basis with respect to this scale.  At any scale, there are two equivalent bases in terms of wavelet functions.  The first basis consists of translates of the father function on the finest scale $J$.  The second basis consists of the father functions on the coarsest scale $j=0$ and mother functions on all intermediate scales $j=0,...,J-1$.  So, for any function we have two equivalent approximations given by
\begin{eqnarray}
f(x) & = & \sum_l a_l \phi_{J,l}(x)\nonumber \\ & = & \sum_l a'_l \phi_{0,l}(x) + \sum_{j=0}^{J-1} \sum_l d_{j,l} \psi_{j,l}(x).
\label{rep}
\end{eqnarray}
It turns out that the first representation is typically dense while the second can often be truncated to a sparse representation by eliminating expansion coefficients with a magnitude below some certain threshold.  This is because the father functions can exactly represent polynomials of degree $K-1$ while the mother functions are orthogonal to such polynomials \cite{daub}. Specifically,
\begin{equation}
\int x^k \psi(x) dx =0, \quad 0 \leq k \leq K-1
\end{equation}
Thus, for any function that is well-represented by low degree polynomials on the scale $J$, most of the coefficients $d_{j,l}$ in the second representation will be small.  These small coefficients can be eliminated with a local error of $O(\epsilon)$, where $\epsilon$ is the threshold of the truncation.  A fast orthogonal transformation known as the discrete wavelet transform \cite{recipe} links the two bases given above.  This allows us to compute projections in the first basis where the single scale and single type of basis function make the approximations accurate and efficient.  Then we can apply the discrete wavelet transform to quickly produce the sparse basis, which is useful for solving linear systems.

\subsection{Application of the Scaling Equation}

Now, we briefly discuss some of the useful results that follow from the scaling equation (\ref{scale}).  For a more detailed treatment see \cite{notes,fewbody}.  First we consider the moments of the father function defined by

\begin{equation}
\left\langle x^k \right\rangle  := \int x^k \phi(x) dx.
\label{moment}
\end{equation}
Applying the scaling equation (\ref{scale}) to (\ref{moment}) gives

\begin{equation}
\left\langle x^k\right\rangle  = \frac{1}{2^k} \sum_l \frac{h_l}{\sqrt{2}} \sum_{m=0}^k {k \choose m} l^{k-m} \left\langle x^m\right\rangle.
\end{equation}

This recursion relation, along with the normalization condition, $\left\langle x^0\right\rangle :=1$, can be used to compute all of the moments of the father function in terms of the filter coefficients, $h_l$.  These moments can be used to construct quadrature rules, which are used to approximate the projection of an arbitrary function, $f(x)$, onto a wavelet basis.  We employ the simplest such quadrature, the one point quadrature \cite{sweldens}.  This quadrature is based on the identity $ \left\langle x^2\right\rangle = \left\langle x\right\rangle^2$ and results in a local error of $O(f^{(3)}(x))$.

It is also important in applications to consider the case where the interval of integration is finite.  Specifically, we consider integrals over left-hand and right-hand endpoints of the form \cite{shann1}

\begin{eqnarray}
 \left\langle x^k\right\rangle_m^+ := \int_0^\infty \phi(x-m) x^k dx,  \nonumber \\ \left\langle x^k\right\rangle_m^- := \int^0_\infty \phi(x-m) x^k dx,
\label{parmom} 
\end{eqnarray}
and
\begin{eqnarray}
\Delta_{mn}^+ \,:= \int_0^\infty \phi(x-m) \phi(x-n) dx, \nonumber \\ \Delta_{mn}^- := \int^0_\infty \phi(x-m) \phi(x-n) dx.
\label{delta}
\end{eqnarray}
Applying the scaling equation (\ref{scale}) to these integrals gives linear relations such as
\begin{equation}
\left\langle x^k\right\rangle_m^+ \,= 2^{-k-1/2} \sum_{l=0}^{2K-1} h_l \left\langle x^k\right\rangle_{2m+l}^+
\end{equation}
and
\begin{equation}
\Delta_{m,n}^+ = \sum_{r=0}^{2K-1} \sum_{s=0}^{2K-1} h_r h_s \Delta_{2m+r,2n+s}^+ .
\end{equation}
These linear systems can be solved for the cases of $m,n=-1,-2,...,-(2K-2)$ using the previously computed moments for $\left\langle x^k \right\rangle_m^+$ and the orthogonality relations for $\Delta_{m,n}^+$.

In \cite{fewbody}, we introduced a method for computing singular integrals of the form
\begin{equation}
S_k := \int \frac{\phi(x-k)}{x+i0^+} dx,
\end{equation}
where $0^+$ is a positive infinitesimal quantity.  Applying the scaling equation (\ref{scale}), gives the degenerate linear relations
\begin{equation}
S_k = \sqrt{2} \sum_{l=0}^{2K-1} h_l S_{2k-l}.
\label{sing}
\end{equation}
These can be supplemented with a normalization condition
\begin{eqnarray}
-i \pi  =  \int_{-a}^a {dx \over x + i 0^+} & = & \nonumber \\ \sum_n \int_{-a}^a \phi (x-n) {dx \over x + i 0^+} & = & \sum_n S_{n:a} ,
\end{eqnarray}
which follows from the identity $1 = \sum_n \phi(x-n)$.  Finally, we need the nonsingular integrals which can be obtained using the recursion relation (\ref{sing}) and the convergent expansion for large $n$ given by

\begin{eqnarray}
S_{n:a} & = & \int_{-a}^a {\phi (x-n) \over x + i0^+}dx    \nonumber \\ & = & \frac{1}{n}  \int_{-a-n}^{a-n} {\phi (y) \over 1+ y/n } dy    \nonumber \\ &=& \frac{1}{n}  \sum_{k=0}^{\infty} \left (\frac{-1}{n} \right )^k \int_{-a-n}^{a-n} \phi (y) y^k dy,
\end{eqnarray}
where the final integrals can be calculated using the methods for equations (\ref{moment}) and (\ref{parmom}).  The values of the singular integrals are given in Table \ref{sing3}.

\begin{table} 
\caption{ Integrals over singularity}
\begin{tabular}{|l|l|l|}
\hline
$S_{-1}$ &-0.1717835441734 & $- i$ 4.041140804162 \\
$S_{-2}$ &-1.7516314066967 & $+ i$ 1.212142562305 \\
$S_{-3}$ &-0.3025942645356 & $- i$ 0.299291822651 \\
$S_{-4}$ &-0.3076858066180 & $- i$ 0.013302589081 \\
\hline					 
\end{tabular}					 
\label{sing3}
\end{table}

For a more thorough and detailed discussion of these calculations and additional properties of wavelets see \cite{notes,fewbody}.

\section{Two-Body T-Matrix in Momentum Space}

The two-body T-matrix is given by the solution to the Lippmann-Schwinger equation

\begin{equation}
T=V+VG_0T,
\end{equation}
where $V$ is the two-body potential and $G_0=(E+i\epsilon-H_0)^{-1}$ is the free two-body propagator. In momentum space, this equation becomes
\begin{widetext}
\begin{equation}
T(p',p,x') = \frac{1}{2\pi} v(p',p,x',1) - m \int_0^\infty dp'' p''^2 \int_{-1}^1 dx'' v(p',p'',x',x'') \frac{1}{p''^2-p_0^2+i\epsilon} T(p'',p,x''),
\label{tmat}
\end{equation}
where $m$ is the mass of the particles, $p_0$ is the on-shell momentum, $x'=\mathbf{\hat{p}'} \cdot \mathbf{\hat{p}},$ $x''=\mathbf{\hat{p}''} \cdot \mathbf{\hat{p}}$, and $v$ is the two-body potential with the azimuthal angle dependence integrated out.  For our calculations, we use a Malfliet-Tjon III potential \cite{pot} with attractive and repulsive parts.  In this case, the azimuthal integration can be carried out analytically giving
\begin{eqnarray}
v(p',p,x',x) & = &\frac{1}{\pi} \left[ \frac{\lambda_{\mathrm{R}}}{\sqrt{(p'^2+p^2-2p'px'x+\mu_{\mathrm{R}})^2-4p'^2p^2(1-x'^2)(1-x^2)}} \right. \nonumber \\ & & \left. -\frac{\lambda_{\mathrm{A}}}{\sqrt{(p'^2+p^2-2p'px'x+\mu_{\mathrm{A}})^2-4p'^2p^2(1-x'^2)(1-x^2)}}  \right] .
\label{pot}
\end{eqnarray}
\end{widetext}

The parameters for this potential are: $\lambda_{\mathrm{A}} =$ -626.8932 MeV fm, $\mu_{\mathrm{A}} =$ 1.55 fm$^{-1}$, $\lambda_{\mathrm{R}} =$ 1438.723 MeV fm,  $\mu_{\mathrm{R}} =$  3.11 fm$^{-1}$, which correspond to those used in \cite{char1}.  We use a nucleon mass such that $1/m =$ 41.47 MeV fm$^2$.

In our work, we consider solutions for the half off-shell T-matrix, $T(p',p_0,x')$.  Traditionally, the T-matrix is decomposed in a partial wave basis using

\begin{equation}
T(p',p_0,x')=\sum_{l=0}^{\infty} \frac{2l+1}{4\pi} T_l(p') P_l(x')
\end{equation}
where the $P_l$ are Legendre polynomials.  Each amplitude $T_l(p')$ must be solved for individually.  For high energies, a significant number of amplitudes may need to be included to ensure convergence \cite{char1}.

The magnitude squared of the on-shell T-matrix is proportional to the differential cross section.  Furthermore, the on-shell partial wave amplitudes, $T_l(p_o)$, can be parameterized as

\begin{equation}
T_l(p_0)=\frac{-2}{\pi} \frac{1}{m p_0} e^{i\delta_l(p_0)} \sin (\delta_l(p_0)),
\label{phase}
\end{equation}
where the $\delta_l(p_0)$ are experimentally determined phase-shifts.  These phase-shifts are used to fit realistic nucleon-nucleon potentials and should be accurately reproduced by any viable solution method.

\section{Wavelet Representation}

To solve equation (\ref{tmat}) we need to transform the half-interval, $[0,\infty)$, corresponding to the momentum variable into a finite interval, $[-a,b]$.  For computational convenience we also transform the interval, $[-1,1]$, associated with the angular variable into the region, $[-c,d]$.  For the first transformation we use the following map

\begin{equation}
p(k):=p_0 \frac{b}{a} \frac{a+k}{b-k}, \quad k(p):=\frac{ab(p-p_0)}{ap+p_0b},
\end{equation}
which maps the scattering singularity at $p''=p_0$ to the origin.  Then we have
\begin{equation}
dp=p_0\frac{b}{a} \frac{(b+a)}{(b-k)^2} dk
\label{dp}
\end{equation}
and
\begin{equation}
\frac{1}{p-p_0} = \frac{a(b-k)}{(a+b)p_0} \frac{1}{k}.
\label{pv}
\end{equation}
The second mapping is the simple linear transformation

\begin{equation}
x(u) := \frac{2u-d+c}{d+c}, \quad  u(x):=\frac{(d+c)x+(d-c)}{2},
\end{equation}
which gives
\begin{equation}
dx = \frac{2}{d+c} du.
\label{dx}
\end{equation}

We now apply these maps to equation (\ref{tmat}) to obtain an equivalent integral equation on the rectangular region $[-a,b] \times [-c,d]$.  For notational convenience we define

\begin{eqnarray}
f(p',x'):=T(p',p_0,x'), \nonumber \\ g(p',x'):=\frac{1}{\pi}v(p',p_0,x',1),
\end{eqnarray}
and for the non-singular part of the kernel
\begin{equation}
L(p',p'',x',x''):=m\frac{v(p',p'',x',x'')p''^2}{p''+ p_0}.
\end{equation}
Now, we let
\begin{eqnarray}
\tilde{f}(k',u'):= f(p(k'),x(u')), \nonumber \\ \tilde{g}(k',u'):=g(p(k'),x(u')),\,
\label{ftilde}
\end{eqnarray}
and
\begin{eqnarray}
& \tilde{L}(k',k'',u',u''):= & \nonumber \\ & L(p(k'),p(k''),x(u'),x(u'')) {\displaystyle \frac{2}{d+c}\frac{b}{b-k''}}. &
\label{ltilde}
\end{eqnarray}
The last factor in this equation comes from applying equations (\ref{dp}), (\ref{pv}) and (\ref{dx}), which gives

\begin{equation}
\frac{1}{p''-p_0} dp'' dx'' = \frac{1}{k''}\frac{2}{d+c}\frac{b}{b-k''} dk'' du''.
\end{equation}
Finally, substituting equations (\ref{ftilde}) and (\ref{ltilde}) into equation (\ref{tmat}) gives

\begin{eqnarray}
& \tilde{f}(k',u')  =  \tilde{g}(k',u') & \nonumber \\ & - {\displaystyle \int_{-a}^b \!\!\!  dk'' \!\!\int_{-c}^d \!\!\! du''  \frac{\tilde{L}(k',k'',u',u'')}{k''}} \tilde{f}(k'',u'') &
\label{trant}
\end{eqnarray}

Now, we project this equation onto the wavelet basis which results in a Galerkin type procedure.  In general, one can choose a separate fine scale in each variable. For notational simplicity, we will consider the case where $J_k=J_u=J$.  In this case, we approximate $\tilde{f}$ using

\begin{equation}
\tilde{f}(k',u') \approx \sum_{m,n} \tilde{f}_{m,n} \phi_{J,m}(k') \phi_{J,n}(u').
\end{equation}
Substituting this in (\ref{trant}) and multiplying by $\phi_{J,m'}(k') \phi_{J,n'}(u')$ and integrating over $k'$ and $u'$ gives the linear equation

\begin{widetext}
\begin{eqnarray}
& \raisebox{2 pt}{$\displaystyle{\sum\limits_{m,n}}$} \, N_{m',n';m,n} \tilde{f}_{m,n}  =  \tilde{g}_{m',n'} & \nonumber \\   - &  \raisebox{2 pt}{$\displaystyle{\sum\limits_{m,n}}$} \, \displaystyle{\int_{-a}^b  \!\!\! dk' \!\!\int_{-c}^d \!\!\! du'  \!\!\int_{-a}^b \!\!\! dk'' \!\!\int_{-c}^d \!\!\! du'' }\phi_{J,m'}(k') \phi_{J,n'}(u')\displaystyle{ \frac{\tilde{L}(k',k'',u',u'')}{k''}} \phi_{J,m}(k'') \phi_{J,n}(u'') \tilde{f}_{m,n}, &
\label{lineq}
\end{eqnarray}
where
\begin{equation}
\tilde{g}_{m',n'} :=   \int_{-a}^b  \!\!\! dk' \!\!\int_{-c}^d \!\!\! du'  \tilde{g}(k',u') \phi_{J,m'}(k') \phi_{J,n'}(u') 
\end{equation}
and
\begin{equation}
 N_{m',n';m,n} := \int_{-a}^b  \!\!\! dk' \!\!\int_{-c}^d \!\!\! du' \phi_{J,m'}(k') \phi_{J,n'}(u') \phi_{J,m}(k') \phi_{J,n}(u').
\end{equation}
We can evaluate $\tilde{g}_{m',n'}$ using the one-point quadrature \cite{sweldens} discussed earlier and an endpoint quadrature based on the partial moments \cite{fewbody}.   $N_{m',n';m,n}$ is simply the direct product of block diagonal matrices consisting of identity blocks and blocks of the form $\Delta^\pm$ given in equation (\ref{delta}).  The final term in equation (\ref{lineq}) can be evaluated using the subtraction
\begin{eqnarray}
\tilde{L}_{m',n';m,n} & := & \int_{-a}^b  \!\!\! dk' \!\!\int_{-c}^d \!\!\! du'  \!\!\int_{-a}^b \!\!\! dk''\!\!\int_{-c}^d \!\!\! du'' \phi_{J,m'}(k') \phi_{J,n'}(u') \frac{\tilde{L}(k',k'',u',u'')}{k''} \phi_{J,m}(k'') \phi_{J,n}(u'') \nonumber \\ & = &  \int_{-a}^b  \!\!\! dk' \!\!\int_{-c}^d \!\!\! du' \!\!\int_{-a}^b \!\!\! dk'' \!\!\int_{-c}^d \!\!\! du''\phi_{J,m'}(k') \phi_{J,n'}(u') \frac{\tilde{L}(k',k'',u',u'')-\tilde{L}(k',0,u',u'')}{k''} \phi_{J,m}(k'') \phi_{J,n}(u'') \nonumber \\ & & + \int_{-a}^b  \!\!\! dk' \!\!\int_{-c}^d \!\!\! du' \!\!\int_{-c}^d \!\!\! du'' \phi_{J,m'}(k') \phi_{J,n'}(u') \tilde{L}(k',0,u',u'')\phi_{J,n}(u'') \int_{-a}^b \frac{\phi_{J,m}(k'')}{k''}   dk''.
\label{subtract}
\end{eqnarray}
\end{widetext}

The first term in this equation is nonsingular and can be approximated using the quadrature methods previously discussed.  Likewise, the $k', u'', u'$ integrations in the second term can be carried out in the same manner.  The final integration over $k''$ can be accomplished using the method following equation (\ref{sing}).

Thus, the problem is reduced to solving a linear system of the form

\begin{equation}
\sum_{m,n} (N_{m',n';m,n}+\tilde{L}_{m',n';m,n}) \tilde{f}_{m,n} = \tilde{g}_{m',n'}.
\label{linsys}
\end{equation}
Once we have solved this equation for $\tilde{f}_{m,n}$ we can substitute this approximate solution back into the right hand side of the original equation to obtain a refined solution.

\section{Wavelet Transform and Sparse Solution}

The eigenvalues of $\Delta^+$ accumulate at $0$ while those of $\Delta^-$ accumulate at $1$ as $K$ increases \cite{jlab}.  This makes the matrix $\mathbf{N}$, and consequently the right hand side of (\ref{linsys}), numerically ill-behaved. To circumvent this difficulty we can precondition the system by inverting $\mathbf{N}$, which is easily accomplished by inverting the two blocks $\Delta^+$ and $\Delta^-$ and using the direct product structure of $\mathbf{N}$.  If we define

\begin{equation}
\mathbf{h}=\mathbf{N\tilde{f}},
\end{equation}
then equation (\ref{linsys}) becomes

\begin{equation}
(\mathbf{I} + \mathbf{LN^{-1}})\mathbf{h} = \mathbf{\tilde{g}}.
\label{consys}
\end{equation}
If we define

\begin{equation}
\mathbf{A}=(\mathbf{I} + \mathbf{LN^{-1}}).
\end{equation}
Then equation (\ref{consys}) is a simply linear system of the form
\begin{equation}
\mathbf{Ah}=\mathbf{\tilde{g}}.
\label{simsys}
\end{equation}
This is a large dense linear system.  However, as shown in equation (\ref{rep}), there are two equivalent representations that are linked by a fast orthogonal transformation.  In two variables, the matrix representation of this transformation is simply the direct product of the one-dimensional transformation matrices that are given in many standard references \cite{recipe}.  If we denote this matrix as $\mathbf{W}$ then we can transform equation (\ref{simsys}) as

\begin{equation}
(\mathbf{WAW^T})\mathbf{Wh} = \mathbf{W\tilde{g}}.
\label{wsys}
\end{equation}
Now we make the definitions

\begin{equation}
\mathbf{\hat{A}}=\mathbf{WAW^T}, \quad \mathbf{\hat{h}}=\mathbf{Wh},  \quad \mathbf{\hat{g}} = \mathbf{W\tilde{g}}.
\end{equation}

Then as mentioned in reference to equation (\ref{rep}) we can truncate the matrix $\mathbf{\hat{A}}$ by eliminating all elements with a magnitude below some certain threshold $\epsilon$, where the error introduced is proportional to $\epsilon$.  The matrix $\mathbf{\hat{A}}$ can be stored in a sparse format such as compressed column format (CCS) \cite{sparse}, which permits both efficient storage and matrix multiplication.  These savings help eliminate computationally costly writing and reading from the hard disk when solving the linear system.

To solve the sparse system we use the complex biconjugate gradient method \cite{recipe,bicon} which we present for general complex matrices in Appendix \ref{biappend}.  This method is a simple and effective iterative method for general sparse matrices.  In addition, this method, like all iterative techniques, is readily amenable to parallel processing of the matrix multiplication.  Once the solution to (\ref{wsys}) is found, it is a simple matter to recover $\mathbf{\tilde{f}}$ by applying the inverse transform $\mathbf{W^T}$ and the inverse matrix $\mathbf{N}^{-1}$.  In particular,

\begin{equation}
\mathbf{\tilde{f}}=\mathbf{N^{-1}W^T\hat{h}}.
\end{equation}

\section{Results and Analysis}

We made calculations using the Daubechies-3 wavelets at scales up to $J=5$ in each variable.  The total number of wavelet basis functions in each variable was taken to be $2^M$, where $M=J+2$.  If we take $a=c=1$, then $b$ and $d$ are determined by

\begin{eqnarray}
b=2^{-J_k} (2^{M_k}-(2K-2)) - a, \nonumber \\ d=2^{-J_u} (2^{M_u}-(2K-2)) - c.
\end{eqnarray}

\begin{figure}[h]
\begin{center}
\rotatebox{00}{\resizebox{3.5in}{!}{
\includegraphics{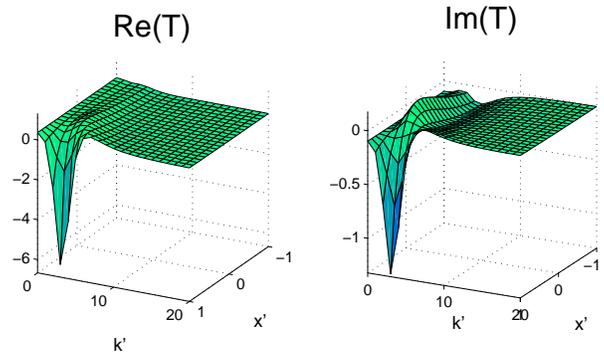}}
}
\caption{(Color online) Half off-shell T-matrix at 800 MeV }
\label{off800}
\end{center}
\end{figure}

Using these parameters calculations were performed at lab energies of 300 and 800 MeV to test the efficacy of the method in different energy regimes.  Figure \ref{off800} shows the real and imaginary parts of the half off-shell T-matrix as a function of momentum, $p'$, and the scattering angle, $x'=\cos(\theta)$ at a scattering energy of  800 MeV.  Daubechies-3 wavelets were used with $M_k=M_u=5$.

\begin{figure}[h]
\begin{center}
\rotatebox{00}{\resizebox{3.5in}{!}{
\includegraphics{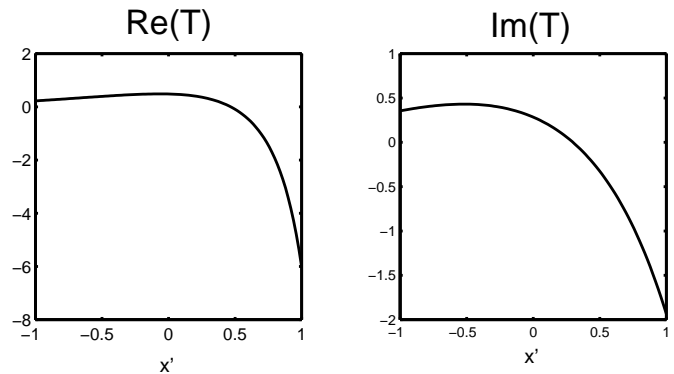}}
}
\caption{On-shell T-matrix at 300 MeV}
\label{on300}
\end{center}
\end{figure}

It can be seen that the real part of the T-matrix is relatively smooth, while the imaginary part does have some structure.  The fact that the T-matrix is smooth with isolated structure suggests that wavelet methods should be able to efficiently compress the matrix $\mathbf{\hat{A}}$.  Figure \ref{on300} shows the on-shell T-matrix as a function of angle, $x'=\cos(\theta)$, at a scattering energy of 300 MeV.  These calculations were made using Daubechies-3 wavelets with $M_k=M_u=5$.
From these graphs the general smoothness of the on-shell amplitude is apparent.  We can also see the forward peaking of the scattering amplitude that is expected at higher energies.

From a numerical standpoint, the first aspect of the calculation to consider is the general convergence of the method as the number of basis functions is increased. Tables \ref{con300} and \ref{con800}  illustrate the convergence of the method as the number of basis functions is increased. The quoted values are for on-shell scattering at an angle of $90^\circ$ using Daubechies-3 wavelets with no truncation.  From Table \ref{con300}, we see that the majority of improvement occurs as $M_k$ is increased.  This can be attributed to the fact that the integral over $k''$ in the kernel is singular and thus requires more basis functions to accurately represent the dependence on this variable.

\begin{table} [hbt]
\caption{Convergence as a function of total number of basis functions: 300 MeV}
\begin{tabular}{lllll}
\hline
$M_k$ & $M_u$ & $\mathrm{Re}(T(p_0,p_0,0))$ && $\mathrm{Im}(T(p_0,p_0,0))$ \\
\hline
 4 & 4 &   0.484065410  && 0.292438234  \\ 
 4 & 5 &   0.483906981  && 0.293504057  \\
 5 & 4 &   0.491111143  && 0.286418162  \\
 5 & 5 &   0.490972783  && 0.287464852  \\
 5 & 6 &   0.490891484  && 0.287452418  \\
 6 & 5 &   0.491773044  && 0.286276262  \\
 6 & 6 &   0.491691220  && 0.286263888  \\
 6 & 7 &   0.491678773  && 0.286256958  \\
 7 & 6 &   0.491772680  && 0.286123199  \\
 7 & 7 &   0.491760404  && 0.286116271  \\
\hline
\end{tabular}
\label{con300}
\end{table}

\begin{table} [hbt]
\caption{Convergence as a function of total number of basis functions: 800 MeV}
\begin{tabular}{lllll}
\hline
$M_k$ & $M_u$ & $\mathrm{Re}(T(p_0,p_0,0))$ && $\mathrm{Im}(T(p_0,p_0,0))$ \\
\hline
 4 & 4 &  0.456127838  && 0.126626540  \\ 
 4 & 5 &  0.454750689  && 0.126515462  \\
 5 & 4 &  0.456227507  && 0.113862313  \\
 5 & 5 &  0.455006434  && 0.113967425  \\
 5 & 6 &  0.454697067  && 0.113367584  \\
 6 & 5 &  0.455242066  && 0.111684819  \\
 6 & 6 &  0.454931562  && 0.111107815  \\
 6 & 7 &  0.454884387  && 0.111005571  \\
 7 & 6 &  0.454978565  && 0.110889988  \\
 7 & 7 &  0.454931334  && 0.110788571  \\
\hline
\end{tabular}
\label{con800}
\end{table}

All of these calculations were made by solving the linear system (\ref{consys}).  It is instructive to consider the behavior if one attempts to solve (\ref{linsys}) using iterative methods without directly inverting $\mathbf{N}$ first.  Table \ref{precon} compares the error in the residual, $\mathrm{e}_n=\left\|\mathbf{\hat{r}}_n\right\|=\left\|\mathbf{\hat{g}}-\mathbf{\hat{A}}\mathbf{\hat{h}}_n\right\|$, as a function of the number of iterations.  As the number of iterations, $n$, is increased the preconditioned method converges very rapidly, while the non-preconditioned method fails to converge adequately.

\begin{table} [b]
\caption{Convergence of the bi-conjugate gradient method}
\begin{tabular}{llllll}
\hline
$n$  & & & &  $\mathrm{e}_n$ (non-preconditioned) & $\mathrm{e}_n$ (preconditioned) \\
\hline
 10 & & & &  3.25$\times 10^{-4}$ & 1.98$\times 10^{-3}$ \\ 
 20 & & & &  4.21$\times 10^{-4}$ & 3.07$\times 10^{-5}$ \\
 30 & & & &  7.23$\times 10^{-5}$ & 2.03$\times 10^{-6}$ \\
 40 & & & &  9.94$\times 10^{-4}$ & 9.10$\times 10^{-10}$ \\
 50 & & & &  1.01$\times 10^{-4}$ & 8.09$\times 10^{-13}$ \\
\hline
\end{tabular}
\label{precon}
\end{table}

\begin{figure}[t]
\begin{center}
\rotatebox{00}{\resizebox{3.5in}{!}{
\includegraphics{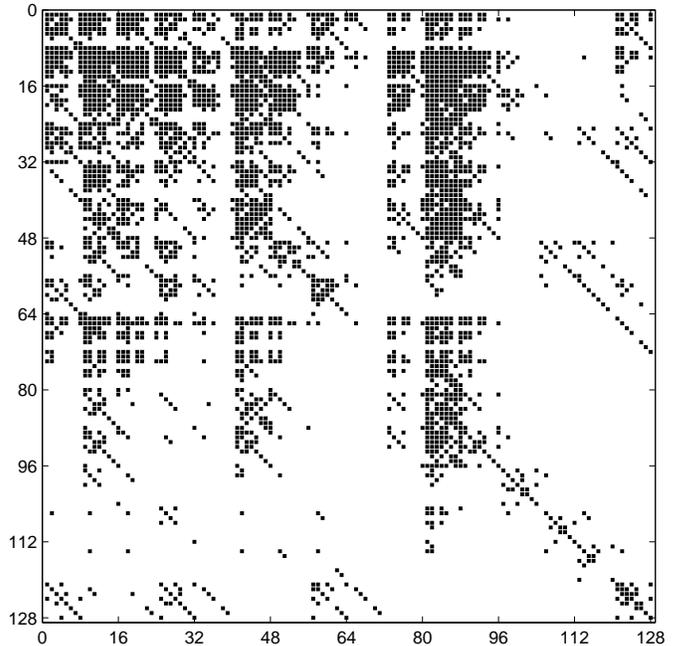}}
}
\caption{Location of the nonzero of elements of $\mathbf{\hat{A}}$ }
\label{spmat}
\end{center}
\end{figure}

Now we turn our attention to the compression of the sparse matrix and its subsequent effect on the calculation.  Figure \ref{spmat} displays a such a representation for scattering at 800 MeV using Daubechies-3 wavelets with $M_k=4$, $M_u=3$.  The plot shows the location of the nonzero elements of $\mathbf{\hat{A}}$ after it has been truncated at the threshold level $\epsilon = 10^{-5}$. This threshold produces a matrix with 19\% of the elements of the full matrix.  The ordering scheme for $\mathbf{\hat{A}}$ used in the plot places the elements associated with finer scales at higher indices.  The degree of sparsity increases considerably as the scale increases, which demonstrates that less and less elements are needed at finer scales.  For the range of test cases we considered, the solution of the sparse linear systems only took 5--10\% of the computational time.  The majority of the time was spent calculating and filtering the dense matrices.

In Table \ref{threshold}, the effect of truncating the matrix $\mathbf{\hat{A}}$ on the convergence of the solution is illustrated for various threshold levels with a lab energy of  800 MeV. This calculation was performed using the Daubechies-3 wavelets with $M_k=M_u=6$. Comparing these results with those in Table \ref{con300}, we see that even keeping just one percent of the matrix elements we are able to reproduce the T-matrix to the same precision as the accuracy of the untruncated matrix.

\begin{table} [t]
\caption{Effect of truncation on the on-shell T-matrix at 800MeV for scattering at $180^\circ,$ $90^\circ$ and $0^\circ$ corresponding to $T(p_0,p_0,-1),$ $T(p_0,p_0,0)$ and $T(p_0,p_0,+1)$}
\begin{tabular}{llllll}
\hline
$\epsilon$  && \% && $\mathrm{Re}(T(p_0,p_0,-1))$ & $\mathrm{Re}(T(p_0,p_0,-1))$ \\
\hline
 0 &&        100 && 0.249235 & -0.0777091 \\ 
$10^{-8}$ &&  23 && 0.249235 & -0.0777093 \\
$10^{-7}$ &&  14 && 0.249234 & -0.0777116 \\
$10^{-6}$ &&  8  && 0.249217 & -0.0777525 \\
$10^{-5}$ &&  1  && 0.248296 & -0.0770660 \\
\hline
$\epsilon$  && \% && $\mathrm{Re}(T(p_0,p_0,0))$ & $\mathrm{Re}(T(p_0,p_0,0))$ \\
\hline
 0 &&        100  && 0.454932 & 0.111108  \\
$10^{-8}$ &&  23  && 0.454932 & 0.111108  \\
$10^{-7}$ &&  14  && 0.454932 & 0.111108  \\
$10^{-6}$ &&  8   && 0.454941 & 0.111117  \\
$10^{-5}$ &&  1   && 0.454966 & 0.111154  \\
\hline
$\epsilon$  && \% && $\mathrm{Re}(T(p_0,p_0,+1))$ & $\mathrm{Re}(T(p_0,p_0,+1))$ \\
\hline
 0 &&        100  &&  -6.16347 & -1.31548  \\
$10^{-8}$ &&  23  &&  -6.16347 & -1.31548  \\
$10^{-7}$ &&  14  &&  -6.16347 & -1.31548  \\
$10^{-6}$ &&  8   &&  -6.16346 & -1.31548  \\
$10^{-5}$ &&  1   &&  -6.16327 & -1.31559 \\
\hline
\end{tabular}
\label{threshold}
\end{table}

Finally, we consider the accuracy of the phase shifts determined by our momentum vector approach.  To calculate the phase shifts we project our T-matrix onto the partial waves using 

\begin{equation}
T_l(p')=2\pi\int_{-1}^1 P_l(x') T(p',p_0,x') dx'.
\end{equation}
We compute the integrals using 20 Gauss-Legendre points \cite{stegun}.  From the $T_l$ it is straightforward to calculate the phase shifts using equation (\ref{phase}).  Table \ref{shifts} displays these phase shifts (calculated for $E_{\mathrm{Lab}}=800$ MeV using Daubechies-3 wavelets with $M_k=M_u=7$ and truncating all but 2\% of the coefficients) compared with phase shifts calculated using standard partial wave techniques.  The agreement between the two methods is very good for all the phase shifts.

\begin{table} [t]
\caption{Comparison of 800 MeV phase shifts with standard methods}
\begin{tabular}{llllllll}
\hline
$l$ &&&& $\delta_l(p_0)$ (Standard) &&&  $\delta_l(p_0)$ (Wavelet) \\
\hline

0 &&&&  -0.2535 &&& -0.2534 \\
1 &&&&   0.2950 &&&  0.2949 \\
2 &&&&   0.3635 &&&  0.3634 \\
3 &&&&   0.2747 &&&  0.2746 \\
4 &&&&   0.1755 &&&  0.1755 \\
5 &&&&   0.1053 &&&  0.1052 \\
6 &&&&   0.06169 &&&  0.06168 \\
7 &&&&   0.03591 &&& 0.03591 \\
8 &&&&   0.02089 &&&  0.02089 \\
9 &&&&   0.01217 &&&  0.01217 \\
10 &&&&   0.007110 &&&  0.007109 \\
11 &&&&   0.004164 &&&  0.004163 \\
12 &&&&   0.002445 &&&  0.002444 \\
13 &&&&   0.001439 &&&  0.001437 \\

\hline
\end{tabular}
\label{shifts}
\end{table}

\section{Conclusions}

We have shown that it is possible to use wavelets to calculate the two-body scattering matrix in terms of momentum vectors without resorting to partial waves. We were able to accurately reproduce the phase shifts of the Malfliet-Tjon potential.  These calculations lead to sparse matrices, which can be efficiently inverted using standard iterative methods.  Application of a simple preconditioning matrix was shown to be necessary to achieve convergence of the iterative methods.  Traditional methods for solving scattering equations in momentum space typically produce dense matrices that require a large amount of storage and are time consuming to invert.  These are promising results because relativistic scattering equations are naturally formulated in momentum space.  Also, the scattering boundary conditions are most easily treated in momentum space.  Wavelet methods can help treat both of these of problems.

One of the main advantages of wavelet methods over methods such as splines is that the wavelet transform presents a method that automatically determines what basis functions are necessary for a given accuracy.  Unfortunately, this also leads to one of the main drawbacks of this method.  In our procedure, a large dense matrix, $\mathbf{A}$, needs to be produced first and then this is transformed to a sparse matrix.  Most of the computational time is spent constructing and transforming this matrix into a sparse format.  The subsequent solution of the sparse linear system takes relatively little computational effort.

For this specific problem, wavelet methods based on momentum vectors may not be necessary.  The maximum number of partial waves that needs to be included to achieve convergence, $l_{max}=14$ \cite{char1}, is simply too small to gain a computational benefit from using wavelets in the angular variable.  To achieve a computational benefit we should use less basis functions in the angular variable than the maximum number of partial waves.  In the three-body problem or at much higher energies, the number of partial waves that need to be included increases considerably and computational benefits may be gained from employing a momentum vector approach.

\begin{acknowledgments}
This work supported 
in part by the U.S. Department of Energy, under contract DE-FG02-86ER40286.
\end{acknowledgments}

\appendix

\newenvironment{Eqnarray}{\arraycolsep 0.14em\begin{eqnarray}}{\end{eqnarray}}

\section{Complex Biconjugate Gradient Method}
\label{biappend}

The biconjugate gradient method \cite{recipe,bicon} is an iterative technique for solving large
matrix equations of the form
\[
\mathbf{A x = b} \,.
\]
The advantage of this method for large sparse matrices is that it only involves
matrix multiplication by $\mathbf{A}$ and its adjoint, both of which can be accomplished efficiently in a sparse storage format such as CCS \cite{sparse}. The algorithm generates a
sequence of approximate solutions, $\mathbf{x}_k$ with residual $\mathbf{r}_k=\mathbf{b-A x}_k$. One
iterates until the norm of the residual is less than some predetermined value.

This method is traditionally formulated for real matrices, but the extension to complex matrices is straightforward.  Below we present the algorithm for general complex matrices.  For our calculations, we start with the initial approximate solution
\[
\mathbf{ x}_0 =\mathbf{ b }
\]
with the residual
\[
\mathbf{ r}_0 = \mathbf{b - A x}_0  \,.
\]
For the initial values of the bi-residual $\mathbf{\bar r}_0$, the direction vector 
$\mathbf{p}_0$, and bi-direction $\mathbf{\bar p}_0$ we use
\begin{Eqnarray}
\mathbf{ \bar r}_0 &=&\mathbf{ b -A^\dagger x}_0  \nonumber \\
\mathbf{ p}_1 &=& \mathbf{r}_0  \nonumber \\
\mathbf{ \bar p}_1 &=& \mathbf{ \bar r}_0  \,. \nonumber
\end{Eqnarray}
\hspace*{-0.28em}Then we use the recurrence relations
\begin{Eqnarray}
\alpha_k &=&  \frac{\mathbf{\bar r}_{k-1}^\dagger \mathbf{r}_{k-1}}{\mathbf{\bar p}_k^\dagger\mathbf{ A p}_k }
                 \nonumber \\
\mathbf{ x}_k &=& \mathbf{x}_{k-1}  + \alpha_k \mathbf{p}_k \nonumber \\
\mathbf{ r}_k &=& \mathbf{r}_{k-1} -\alpha_k \mathbf{A p}_k \nonumber \\
\mathbf{ \bar r}_k &=& \mathbf{r}_{k-1}  - \alpha_k^\ast \mathbf{A^\dagger \bar p}_k \nonumber \\
\beta_k &=&  \frac{\mathbf{\bar r}_k^\dagger\mathbf{ r}_k}{\mathbf{\bar r}_{k-1}^\dagger \mathbf{r}_{k-1}} 
                 \nonumber \\
\mathbf{ p}_{k+1} &=& \mathbf{r}_k  + \beta_k \mathbf{p}_k \nonumber \\
\mathbf{ \bar p}_{k+1} &=&\mathbf{ \bar r}_k  + \beta_k^\ast \mathbf{\bar p}_k \,. \nonumber
\end{Eqnarray}
\hspace*{-0.28em}to generate an improved approximation. This is repeated until
the desired accuracy is obtained.  We measure the accuracy by the $\ell^2(\mathbb{C})$ norm of the residual,
\[
\mathrm{e}_k=\left\|\mathbf{\hat{r}}_k\right\|=\sqrt{\mathbf{r^\dagger r}} \, .\nonumber
\]

\end{document}